\documentclass[mathleft]{an}
\usepackage{graphicx}
\usepackage{times}
\def\4u{4U~2206+54}
 
\def\kms{km$\,$s$^{-\!1}$}

\def\msun{M$_{\odot}$}
\newcommand{\ltsima} {$\; \buildrel < \over \sim \;$} 
\newcommand{\simlt} {\lower.5ex\hbox{\ltsima}} % < over MMM 
\newcommand{\gtsima} {$\; \buildrel > \over \sim \;$} 
\newcommand{\simgt} {\lower.5ex\hbox{\gtsima}} % > over MMM 

\begin{document}

% The following seven commands are intended for editorial usage and should be ignored by
% the author(s).
\Pagespan{789}{}% Document's page range. 
% If second parameter is left empty, the last page is computed automatically.
\Yearpublication{2014}%
\Yearsubmission{2014}%
\Month{11}%   
\Volume{999}%  
\Issue{88}% 
% \DOI{This.is/not.aDOI}% 

\title{Orbital Parameters of the High-Mass X-ray Binary 4U~2206+54\thanks{based on observations obtained with 2m RCC telescope at Rozhen NAO, Bulgaria}}

\author{K. A. Stoyanov\inst{1}\fnmsep\thanks{Corresponding author:
  \email{kstoyanov@astro.bas.bg}\newline}
%Example 
%for footnote, note the usage of the \texttt{fnmsep}
%command as separator between institute number and footnote mark} 
\and R. K. Zamanov\inst{1}
\and G. Y. Latev\inst{1}
\and A. Y. Abedin\inst{2}
\and N. A. Tomov\inst{1}
}
\titlerunning{Orbital Parameters of \4u}
\authorrunning{Stoyanov, Zamanov, Latev, Abedin \& Tomov}
\institute{Institute of Astronomy and National Astronomical Observatory, Bulgarian Academy of Sciences, 
       72 Tsarighradsko Shousse Blvd., 1784 Sofia, Bulgaria
       \and
       Department of Physics and Astronomy, The University of Western Ontario, London, ON, N6A 3K7, Canada}

\received{2014 Apr 30}
\accepted{2014 Aug 28}
\publonline{2014 Dec 01}


\abstract{
We present new radial velocities of the high-mass X-ray binary star \4u based on optical spectra obtained with the
Coud\'e spectrograph at the 2m RCC telescope at the Rozhen National Astronomical Observatory, Bulgaria in the period
 November 2011 -- July 2013 .
The radial velocity curve of the HeI $\lambda$6678 \AA\ line is modeled with an orbital period P$_{orb}$ = 9.568~d and 
an eccentricity of $e$ = 0.3. These new
measurements of the radial velocity resolve the disagreements of the orbital period discussions.
}
\maketitle

\section{Introduction} 
\4u (BD~53$^0$2790; LS~III~+54~16) is a persistent high-mass X-ray binary star at a distance of 2.6~kpc (Blay et al. 2006).
It was first detected as an X-ray source by the $\it{Uhuru}$ satellite (Giacconi et al. 1972).
The system also appeared in the Ariel~V catalogue (Warwick et al. 1981).
The mass donor is classified as an O9.5Vp star with a higher than normal helium abundance,
underfilling its Roche lobe and losing mass 
via a slow but dense stellar wind, $\upsilon \sim$ 350 \kms\ (Rib{\'o} et al. 2006). 
However, there are some metallic lines typical for a later-type spectrum (Negueruela \& Reig 2001).
The compact object is a neutron star (Torrej{\'o}n et al. 2004) with spin period
P$_s$ = 5554$\pm$9~s (Finger et al. 2010). The X-ray spectrum of \4u is typical for a neutron star accreting material onto its
magnetic poles.\\
The orbital period of the system is a subject of discussion. 
X-ray monitoring by RXTE has suggested an orbital period of 9.568
(Corbet \& Peele 2001), but later SWIFT/BAT and RXTE observations found a modulation of about 19.25~d, twice of the 9.6-day period
(Corbet, Markwardt \& Tueller 2007). The combination of slow X-ray pulsations (Reig et al. 2009; Finger et al. 2010) and the
spin-down rate of the neutron star (Reig, Torrej{\'o}n \& Blay 2012) suggests a strong magnetic field $\sim 10^{14}$~G.
If the high magnetic field is confirmed, then \4u will become the first kind of an accreting magnetar.

Here we present new optical spectroscopy and radial velocity measurements of \4u. 

\section{Observations and data reduction}
All  data reported here are obtained
by the 2m RCC  telescope of the Rozhen National Astronomical Observatory, Bulgaria. 
\4u is observed  between November 2011 and July 2013 with the Coud\'e spectrograph of the telescope.
We used two gratings with resolutions of R = 30~000 and R = 15~000, respectively. The log of observations is given in Table \ref{log}.

The spectra are reduced in the standard way including bias removal, 
flat-field correction, wavelength calibration and correction for the Earth's motion. 
Pre-processings and measurements of the radial velocities are performed using standard routines provided by 
IRAF\footnote{IRAF is distributed by the National Optical Astronomy Observatory, which is operated by the Association of Universities for Research in Astronomy, Inc., under cooperative agreement with the National Sciences Foundation}. 
The spectra obtained within each observational night are processed and measured independently. 

Our spectra include two prominent spectral features - H$\alpha$ and HeI $\lambda$6678 \AA. 
The double-peaked H$\alpha$ line is dominated by emission from the circumstellar disk of the donor star. 
%While the absorbtion trough does follow the orbital motion of the star, 
We did not use H$\alpha$ line to measure the
orbit since its radial velocity measurements may be strongly affected by changes in the disk structure.
The HeI $\lambda$6678 line is less affected by changes in the disk, therefore we use only this line for radial velocity
measurements.

\begin{figure}
 \vspace{8.0 cm}  
 \includegraphics{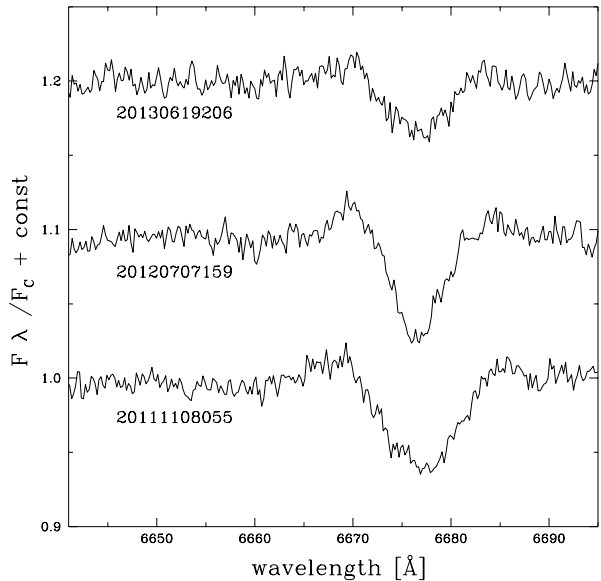}   
 \caption[]{Examples of HeI $\lambda$6678 \AA\ line. Below the spectra the individual IDs are given (see Table \ref{log}).}
    \label{profile}
\end{figure}

Three examples of HeI $\lambda$6678 line are plotted in Fig.\ref{profile}.
\begin{table*}
\caption{Observations of \4u. Given here are as follows: the ID of the spectrum, MJD of the start of the exposure, 
the exposure time, the resolution of the grating (R30 for R=30~000 and R15 for R=15~000), 
the orbital phase folded with P$_{orb}$ = 9.568~d, the radial velocity of the HeI $\lambda$6678 line,
and (O - C) errors.}
\label{log}      
\centering                          
\begin{tabular}{lccccccrccc}       
\hline\hline                 
 &  &  \\                
spectrum ID  &MJD -start& Exp Time & R & Orbital &  V$_r$HeI    &  (O - C)   \\ 	
yyyymmddxxx   &         &  [min] &	&  Phase   & [\kms]    &   [\kms]    \\
\\	          
\hline           												       
\\
20111108055  &	2455873.896963   & 20	& R15  &   0.272     &      -24.8  & -1.8 & \\ 
20111108056  &	2455873.912325   & 20	& R15  &   0.274     &      -20.3  &  2.9 & \\
20120706124  &	2456114.974316   & 20	& R15  &   0.469     &      -71.8  & -1.5 & \\
20120706125  &  2456114.989071   & 20	& R15  &   0.470     &      -79.8  & -9.6 & \\
20120707159  &	2456115.983374   & 20	& R15  &   0.574     &      -94.2  & -16.5 & \\
20120707160  &	2456115.997523   & 20	& R15  &   0.576     &      -84.3  & -6.6 & \\
20120708024  &  2456116.885644   & 20	& R15  &   0.668     &      -89.3  & -13.8 & \\
20120708025  &	2456116.900287   & 20	& R15  &   0.670     &      -81.6  & -6.6 & \\
20120709092  &	2456117.905134   & 30	& R15  &   0.775     &      -88.3  & -19.4 & \\
20120709098  &	2456117.955419   & 30   & R30  &   0.780     &      -80.2  & -11.7 & \\
20120710130  &	2456118.891829   & 30   & R30  &   0.878     &      -81.6  & -21.1 & \\
20120830453  &	2456169.795944   & 30	& R30  &   0.198     &      -27.5  & -2.6 & \\
20120830454  &  2456169.816999   & 30	& R30  &   0.201     &      -34.2  & -9.6 & \\
20120927415  &	2456197.824752   & 20   & R15  &   0.128     &      -54.4  & -21.7 & \\
20120927416  &	2456197.838901   & 20	& R15  &   0.129     &      -41.8  & -9.4 & \\
20121005639  &	2456205.765701   & 30   & R30  &   0.958     &      -56.6  & 4.3 & \\
20121005640  &	2456205.786759   & 30	& R30  &   0.960     &      -46.2  & 16.5 & \\
20121006736  &	2456206.762868   & 30   & R30  &   0.062     &      -51.8  & -11.1 & \\
20121006737  &	2456206.783925   & 30	& R30  &   0.064     &      -41.0  & -0.5 & \\
20121025107  &	2456225.733967   & 30   & R30  &   0.045     &      -28.5  & 14.4 & \\
20121025108  &	2456225.755025   & 30	& R30  &   0.047     &      -29.4  & 13.7 & \\
20121026204  &	2456226.720280   & 30   & R30  &   0.148     &      -30.5  & -0.3 & \\
20121026205  &	2456226.741339   & 30	& R30  &   0.150     &      -27.3  & 2.6 & \\
20121104034  &	2456235.755836   & 20	& R15  &   0.092     &      -37.8  & -0.7 & \\
20121104035  &  2456235.771710   & 20	& R15  &   0.094     &      -35.5  & 1.3 & \\
20130102019  &	2456294.756720   & 20	& R15  &   0.259     &      -9.4   & 13.0 & \\
20130102020  &	2456294.770866   & 20	& R15  &   0.260     &      -8.1   & 14.5 & \\
20130123098  &	2456315.695479   & 20	& R15  &   0.447     &      -50.7  & 15.6 & \\
20130123099  &	2456315.710368   & 20	& R15  &   0.449     &      -55.8  & 10.8 & \\
20130520065  &	2456432.934320   & 20   & R15  &   0.700     &      -58.8  & 15.1 & \\
20130521110  &	2456433.948816   & 20	& R15  &   0.806     &      -64.8  & 1.8 & \\
20130521111  &	2456433.962972   & 20	& R15  &   0.808     &      -61.6  & 4.8 & \\
20130522136  &	2456434.907042   & 30	& R15  &   0.906     &      -42.4  & 15.3 & \\
20130522137  &	2456434.928138   & 30	& R15  &   0.909     &      -44.7  & 12.9 & \\
20130525168  &	2456437.902431   & 30	& R15  &   0.220     &      -23.9  & -0.7 & \\
20130525169  &	2456437.923521   & 30	& R15  &   0.222     &      -25.9  & -2.8 & \\
20130619206  &	2456462.970446   & 20	& R15  &   0.840     &      -60.9  & 2.9 & \\
20130718030  &	2456492.014043   & 20	& R15  &   0.875     &      -51.7  & 8.9 & \\
20130719041  &	2456492.807739   & 20	& R15  &   0.958     &      -48.4  & 4.3 & \\
20130719044  &	2456492.827451   & 20	& R15  &   0.960     &      -35.9  & 6.1 & \\
20130723140  &	2456496.968814   & 20	& R15  &   0.393     &      -65.6  & -13.5 & \\
20130724170  &	2456497.786707   & 20	& R15  &   0.478     &      -64.2  & 7.5 & \\
20130724171  &	2456497.800856   & 20	& R15  &   0.480     &      -64.2  & 7.7 & \\
\hline                                   
\end{tabular}
\label{log}
\end{table*}

\section{Orbital solution}
\begin{figure}
 \vspace{8.0 cm}  
 \includegraphics{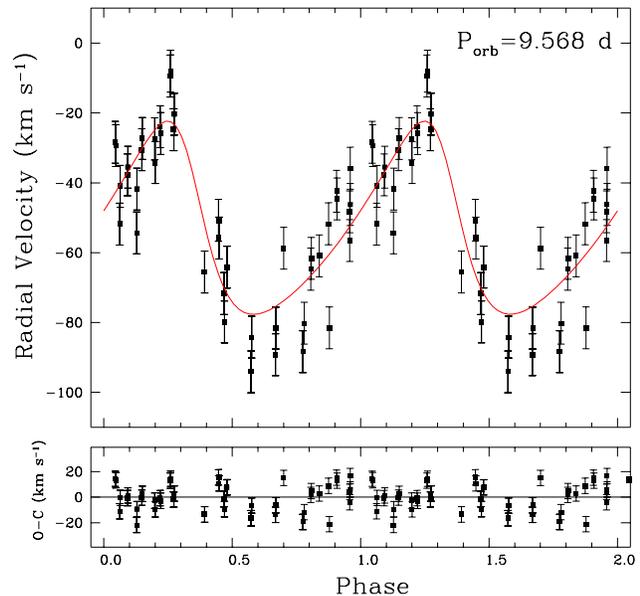}   
 \caption[]{Radial velocity curve of \4u ({\it upper panel}). The best-fitting solution is overplotted with red line.
	    {\it The lower panel} shows the residuals of the fit.}
    \label{resid}
\end{figure}

\begin{figure}
 \vspace{8.0 cm}  
 \includegraphics{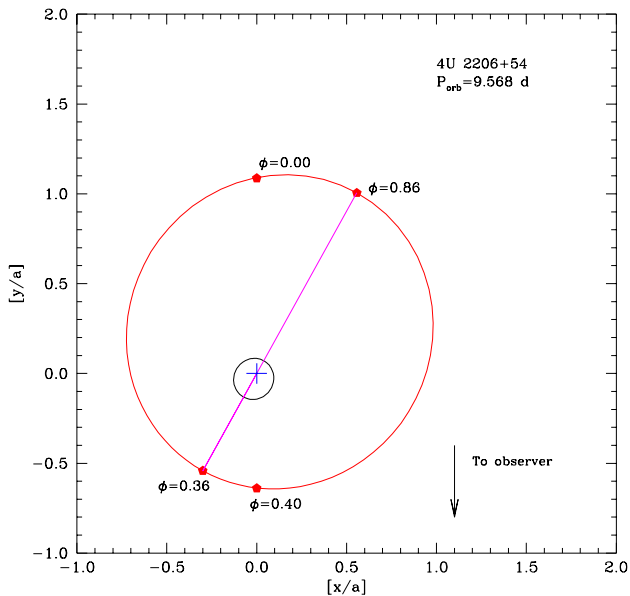}   
 \caption[]{Orbital geometry of \4u showing the relative orbits of the 10.7 \msun\ donor star and the 1.4 \msun\ neutron star.
            The relevant phases of the periastron, apastron, and the conjunctions are marked. The center of the mass is
             indicated with a cross. }
    \label{orbit}
\end{figure}

The radial velocity curve was modeled with an eccentric orbital solution. 
We used the {\sc Phoebe} program (Pr\v{s}a \& Zwitter 2005) to solve the component parameters and independently applied the
{\sc Nightfall} program (Wichmann 2006). Both programs are based on the Wilson~-~Devinney code (Wilson \& Devinney 1971). 
We estimate the errors by comparing the consistency of the output from the two programs. 
Using the both possible orbital periods, we estimate the standard deviation of the fit to reveal the proper
orbital period. P$_{orb}$ = 9.568~d and P$_{orb}$ = 19.25~d give standard deviations of the fits of $\sigma$=10.3~\kms
and $\sigma$=40.8~\kms, respectively. For our orbital solution,
we took P$_{orb}$ = 9.568~d. In Sect.\ref{Orbital} we discuss the reasons for this choice in more details.\\
For the orbital solution, we adjusted the following orbital parameters: eccentricity of the orbit {\it e}, 
periastron passage $\omega$, systemic velocity $\gamma$, time of the conjunction passage MJD$_0$, semiamplitude velocity  K$_1$, projected separation
$a_1 \sin i$ , mass function $f(M)$, and the standard deviation of the fit $\sigma$.
A set of fixed parameters necessary for the fitting procedure is applied to match the properties of the optical companion.
The best-fitting orbital parameters are listed in Table \ref{Orbit}. In Fig.\ref{resid} are plotted the radial velocity curve,
the best-fitting solution, and the residuals of the fit. The geometry of the orbit is illustrated in Fig.\ref{orbit}.

\begin{table}
\centering
\caption[]{Orbital parameters of \4u.}
\label{tabrvfit}
\begin{tabular}{lcc}
\hline
\hline
Parameter                  &                   &                \\
\hline  
P$_{orb}$ (d)		   &   9.568*	       &                 \\     
e                          &   0.30$\pm$0.02   &                \\
$\omega$ (deg)             &    61$^{0}$.2$\pm$1     &                \\
$\gamma$ (km s$^{-1}$)     &   -54.5$\pm$1       &                \\
MJD$_0$ (HJD-2,450,000)     & 5871.67$\pm$0.05       &                \\
$K_1$ (km s$^{-1}$)        &    30.5$\pm$3       &                \\
$a_1 \sin i$ (R$_{\odot}$) &  3.76$\pm$0.05    &                \\
$f(M)$ (M$_{\odot}$)       & 0.0232$\pm$0.0045 &                \\
$\sigma$ (km s$^{-1}$)     &    10.3             &                \\	
\hline
\end{tabular}
\\
$*$for the purposes of our orbital solution, we used this value.
\label{Orbit}
\end{table}

%\begin{equation}
 %f(m) = \frac{m_2^3~sin^{3}~i}{(m_1 + m_2)^2} = (1.0361 \times 10^{-7})~(1 -
 %e^2)^{3/2}~K_1^3~P_{orb}~[M_{\sun}] 
%\end{equation}

\section{Discussion}
\label{Discussion}
\subsection{Orbital period}
\label{Orbital}
The orbital periods of the Be/X-ray binaries are in the range from $\sim$10~d to $\sim$1~year.
For example, SAX~J2103.5-4545 has the shortest known orbital period among the Be/X-ray binaries: 
P$_{orb}$ = 12.7~d (Camero Arranz et al. 2007).

The orbital period of \4u is still not well determined. X-ray observations revealed two possible values for the 
orbital period: P$_{orb}$ = 9.568 $\pm$ 0.004~d and P$_{orb}$ = 19.25~d (Corbet \& Peele 2001; Corbet, Markwardt \& Tueller 2007).
We found P$_{orb}$ = 9.55 $\pm$ 0.05 using the Phase Dispersion Minimization method (Stellingwerf 1978), which is not an improvement over Corbet \& Peele (2001).

\begin{figure}
 \vspace{8.0 cm}  
 \includegraphics{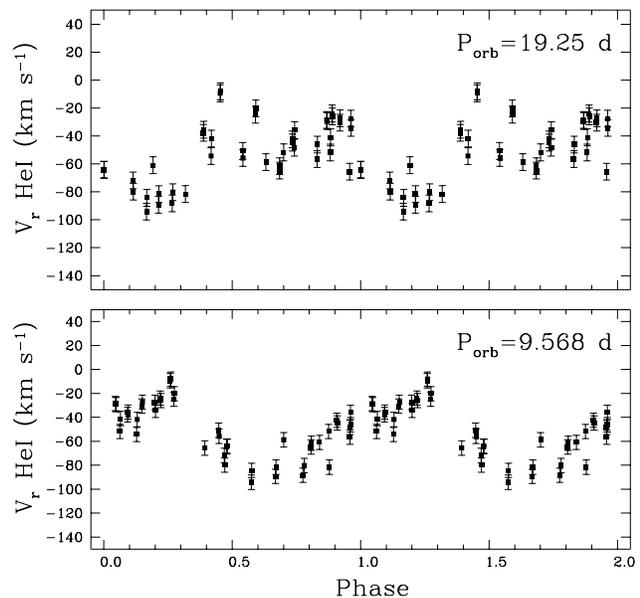}   
 \caption[]{The radial-velocity curve of \4u folded with $P_{orb}$=9.658~d ({\it lower panel})
            and $P_{orb}$=19.25~d ({\it upper panel}). }
    \label{4U.rv}
\end{figure}

The correlation between the equivalent width EW(H$\alpha$) and orbital period for Be/X-ray binaries is explored in details in 
Reig, Fabregat, \& Coe (1997) and Reig (2011). The systems with shorter orbital periods show less emission
in H$\alpha$ than those with longer orbital periods (Fig.15 in Reig (2011)).\\
We have measured the EW(H$\alpha$)
in our spectra in order to find a clue for the real orbital period of \4u. Our minimum and maximum values for 
the EW(H$\alpha$) are 0.51~\AA\ and 3.12~\AA\, respectively. Blay et al. (2006) have measured a
maximum value of EW(H$\alpha$) of 7.3~\AA. According to Fig.15 in Reig (2011), 
the orbital period of the system should be the shorter one: P$_{orb}$ = 9.568~d.\\
Moreover, in Fig.\ref{4U.rv} 
we plot the radial velocities of the HeI $\lambda$6678 line folded with the two possible orbital periods. 
It is clearly visible that shorter orbital period modulate the data better than the longer orbital period. 
If the P$_{orb}$ = 9.568~d is confirmed, \4u will become the Be/X-ray binary with the shortest orbital period.
It will be another addition to the peculiar features that divert the system from the classical Be/X-ray binaries.

\subsection{Orbital eccentricity}

There is a group of Be/X-ray binaries (X~Per, GS~0834-430, KS~1947+300, XTE~J1543-568, and 2S~1553-542) 
characterized by very low eccentricities: $e \leq$ 0.2 (Reig 2011). Their low eccentricity requires that the
compact object received a much lower kick velocity at birth than previously assumed by current evolutionary 
models (Pfahl et al. 2002). These objects have P$_{orb} \geq$ 30~d.

Most of the Be/X-ray binaries have moderately eccentric orbits with $e \geq$ 0.3. For them the tidal force acts
as a decelerator of the rotation of the mass donor in order to reach an equilibrium state, 
i.e. a circular and synchronized orbit (Stoyanov \& Zamanov 2009).

It will be interesting to check whether the rotation of the mass donor is pseudosynchronized with the orbital motion 
of the compact object in the case of massive and short-period system such as \4u.

\section{Conclusions}

On the basis of radial velocity measurements of the HeI $\lambda$6678 \AA\ line, we measured the orbital parameters of the
high-mass X-ray binary star \4u. We found that the orbit of the system should be eccentric with $e$ = 0.3, if the orbital period is
9.568~d. We discussed the probability that 4U 2206+54 is a Galactic Be/X-ray binary with the shortest orbital period known up
today.

\acknowledgements
We thank the anonymous referee for the constructive comments. This work was supported by the OP ``HRD``, 
ESF and Bulgarian Ministry of Education, Youth and Science under the contract 
BG051PO001-3.3.06-0047.

%\newpage 

\end{document}